\documentclass{aa}
\usepackage{natbib}
\bibpunct{(}{)}{;}{a}{}{,} 

\newcommand{\hea}[1]{{#1}}

\usepackage[colorlinks=true,urlcolor=blue,citecolor=blue,linkcolor=blue]{hyperref}
\usepackage{graphicx}
\usepackage{txfonts}

\usepackage[colorlinks=true,urlcolor=blue,citecolor=blue,linkcolor=blue]{hyperref}
\begin{document} 

        \title{Deep solar ALMA neural network estimator for image refinement and estimates of small-scale dynamics}

   \author{Henrik Eklund \inst{1,2,3}
  }
   
  \authorrunning{Eklund {et~al.}}
   \institute{Institute for Solar Physics, Department of Astronomy, Stockholm University AlbaNova University Centre, SE-106 91 Stockholm, Sweden
   \and
   Rosseland Centre for Solar  Physics, University of Oslo, Postboks 1029 Blindern, N-0315 Oslo, Norway
            \and
            Institute of  Theoretical Astrophysics, University of Oslo, Postboks 1029 Blindern, N-0315 Oslo, Norway \\
            \email{henrik.eklund@astro.su.se}
}

\def\corrAuthor{Henrik Eklund}

   \date{Received --- ; accepted --- }

\abstract{
The solar atmosphere is highly dynamic, and observing the small-scale features is valuable for interpretations of the underlying physical processes. 
The contrasts and magnitude of the observable signatures of small-scale features degrade as angular resolution decreases.
}
{
The estimates of the degradation 
associated with the \hea{observational angular resolution} allows a more accurate analysis of the data. 
}
{
High-cadence time-series of synthetic observable maps 
at $\lambda=1.25$~mm
were produced from three-dimensional magnetohydrodynamic Bifrost simulations of the solar atmosphere and degraded to the angular resolution corresponding to observational data with the Atacama Large Millimeter/sub-millimeter Array (ALMA).
The Deep Solar ALMA Neural Network Estimator (Deep-SANNE) is an artificial neural network trained to improve the resolution and contrast of solar observations.
This is done by recognizing 
dynamic patterns in both the spatial and temporal domains of small-scale features at an angular resolution corresponding to observational data and correlated them to highly resolved nondegraded data from the magnetohydrodynamic simulations.
A second simulation, previously never seen by Deep-SANNE, was used to validate the performance.
}
{
Deep-SANNE provides maps of the estimated degradation of the brightness temperature across the field of view, which can be used to filter for locations that \hea{most probably} show a high \hea{accuracy
and} as correction factors 
in order to construct refined images that show higher contrast and more accurate brightness temperatures than at the observational resolution.
Deep-SANNE reveals more small-scale features in the data and achieves a good performance in estimating the excess temperature of brightening events with an average of $94.0\%$ relative to the highly resolved data, compared to $43.7\%$ at the observational resolution.
By using the additional information of the temporal domain, Deep-SANNE can restore high contrasts better than a standard two-dimensional deconvolver technique. 
In addition, Deep-SANNE is applied on observational solar ALMA data, for which it also reveals eventual artifacts that were introduced during the image reconstruction process,  in addition to improving the contrast. It is important to account for eventual artifacts in the analysis.
}
{
The Deep-SANNE estimates and refined images are useful for an analysis of small-scale and dynamic features. They can identify locations in the data with high accuracy for an in-depth analysis and allow a more meaningful interpretation of solar observations.
}


   \maketitle

\section{Introduction}

The solar chromosphere is highly dynamic. Its structures show a large variety of temporal and spatial scales. Observing the small-scale dynamics can often be imperative for understanding the underlying physical processes that take place in the atmosphere. 
However, the ability to detect small-scale features is heavily dependent on the angular resolution of the observations.
The angular resolution increases with the size of the aperture and with decreasing wavelength of the radiation.
At millimeter (mm) wavelengths, the Atacama Large Millimeter/sub-millimeter Array (ALMA) provides large advances in terms of sensitivity and angular resolution, and solar observations are currently offered down to approximately $0.62$~arcsec at $1.25$~mm \citep[receiver band~6;][]{ALMA_Cycle_9_Prop_Guide} and to e.g., a resolution $\sim0.92$~arcsec at $3.0$~mm (receiver band~3).

The mm wavelength radiation is formed from free-free emission under local thermal equilibrium, so the measured intensities depend linearly on the local plasma temperature at the sampled heights \citep[see, e.g.,][and references therein]{2016SSRv..200....1W} and can be expressed as the brightness temperature, which is valuable for studying the transport of energy and heating at different layers of the solar atmosphere.
Dynamic small-scale features and brightening events, such as excited by propagating shocks \citep{2021RSPTA.37900185E}, have been studied in ALMA observations \citep{2020A&A...644A.152E}.
However, the contrast of the features is severely degraded because of the large size of the point spread function 
\citep[PSF; see, e.g.,][]{2015A&A...575A..15L, 2020A&A...644A.152E}, which makes the interpretation of the data and estimation of possible energy dissipation from the shocks challenging.

The angular resolution affects the detection and identification of small-scale features in the spatial domain, but it also affects variations and oscillations in the temporal domain of the data \citep{2021NatAs...5....5J, 2021A&A...656A..68E, 2021RSPTA.37900174J}.
The degree of the degradation of a particular feature is dependent on the angular resolution, but also the distribution and spatial scales of the surrounding intensities. 

A statistical approach detecting brightening events in synthetic observables from a three-dimensional magnetohydrodynamic (MHD) simulation from the code Bifrost \citep{2016A&A...585A...4C, 2011A&A...531A.154G} was deployed by \cite{2021A&A...656A..68E} to derive correction factors of the degradation of their brightness temperature amplitude at each ALMA receiver band at different angular resolutions.
However, these correction factors come with significant uncertainties, which means that different events are differently well resolved.
For instance, bright, unresolved features smaller than the PSF and less bright, and better resolved extended features larger than the PSF appear with similar brightness temperature in observation data.

This motivated the development the Deep Solar ALMA Neural Network Estimator (Deep-SANNE), a deep neural network trained to recognize dynamic spatio-temporal patterns. It estimates the accuracy of the brightness temperatures across the field of view and produces refined images that are corrected for the resolution degradation.
Deep-SANNE can be applied to science-ready high-cadence time-series of solar ALMA data. It serves as a complementary tool for analysis
and can aid in the choice of locations with the highest accuracy of the brightness temperatures for an in-depth analysis of events in the data.

This paper is structured as follows. In Sect.~\ref{sec:methods} the architecture of the artificial neural network and the simulations used to train it and to validate the estimates are presented. 
In Sect.~\ref{sec:results} the resulting estimates of the validation data that were performed by the neural network are presented, together with a few applications of the estimates for analysis. The application of Deep-SANNE on an observational solar ALMA data set is also shown.
In Sect.~\ref{sec:disc} and in Sect.~\ref{sec:conc} we conclude and describe future developments.


\section{Methods}\label{sec:methods}

Among the available techniques, neural networks have shown a very good performance in terms of accuracy and speed in different applications such as pattern detection, radiative transfer calculations, and image reconstruction
\citep[see, e.g.,][]{2018MNRAS.481.5014I, 2019A&A...629A..99D, 2021A&A...648A..53D}.
In a previous study of \cite{2021A&A...656A..68E}, it was concluded that dynamical small-scale signatures have a important imprint in the spatial and temporal domains. The motivation therefore is to use a neural network that combines spatial and temporal analysis modules to improve the determination of the degradation.

\subsection{Neural networks}

When two-dimensional (2D) data such as images are sutdied, convolutional neural networks \citep[CNN; see e.g.,][]{2016arXiv160307285D} are very useful. The convolutional
operations
capture information in the spatial domain. There are a numerous types of convolutional neural networks that have benefits for different applications.
They have already been used in solar physics to improve the spatial resolution of solar observations \citep{2018A&A...614A...5D, 2021MNRAS.501.2647A}, but the temporal information of these observations was not considered.

Recurrent neural networks \citep[RNN; ][]{rumelhart1986learning} are the type of neural networks that are used to handle information in the temporal domain and process sequences of data, where the output is dependent on the previous elements in the data stream. 
These types of neural networks, usually trained by a backpropagation method \citep[see, e.g.,][]{1992..hecht..theory} to calculate the gradient of the loss function \citep[see, e.g.,][]{2017..goodfellow..deep}, may be affected by what is referred to as a vanishing gradient \citep{2001_hochreiter_gradient} and long-term memory loss, which prevents them from learning over longer sequences of data.

Long-short-term memory \citep[LSTM; ][]{1997_hochreiter_long} is a form of a recurrent neural network where a mechanism of different gates is used to regulate the passage and flow of information through the network, which effectively takes care of the vanishing gradient problem.
This ability makes an LSTM architecture suitable for handling long sequences of data and recognizing patterns at short scales of the sequence. It also preserves the context from the beginning of the data stream, hence the name long-short-term memory. 
LSTM neural networks have been shown to excel in several fields of applications, such as speech recognition and handwriting recognition \citep{2008_graves_offline}, where information about the context needs to be preserved through the data sequence.

Basic LSTM architectures are used to study one-dimensional sequences of data, while CNNs are used for images at a given time. In the following, we explain how they can be combined in a single architecture.

\subsection{Deep-SANNE architecture}\label{sect:methods-deep_sanne}
In the current work, we use an artificial deep neural network with an architecture based on convolutional LSTM cells, where convolutional operations are performed at the gates within the LSTM cell, instead of at the regular matrix multiplications. 
The main advantage of the convolutional LSTM is that the state of a certain cell is determined by the past states of the neighboring cells (image pixels) that are within reach of the convolutional kernel, which allows capturing spatial features and studying their dynamics and evolution in time.
Each LSTM cell consists of a number of gates,
input, forget, cell, output, and hidden,
which each have a different task
to dynamically update the different weights
that regulate the flow of information through the neural network\footnote{Keras, \url{https://keras.io/}} \citep{2015arXiv150604214S}.

\begin{figure*}[tp!]
\centering
\sidecaption
\includegraphics[width=12cm]{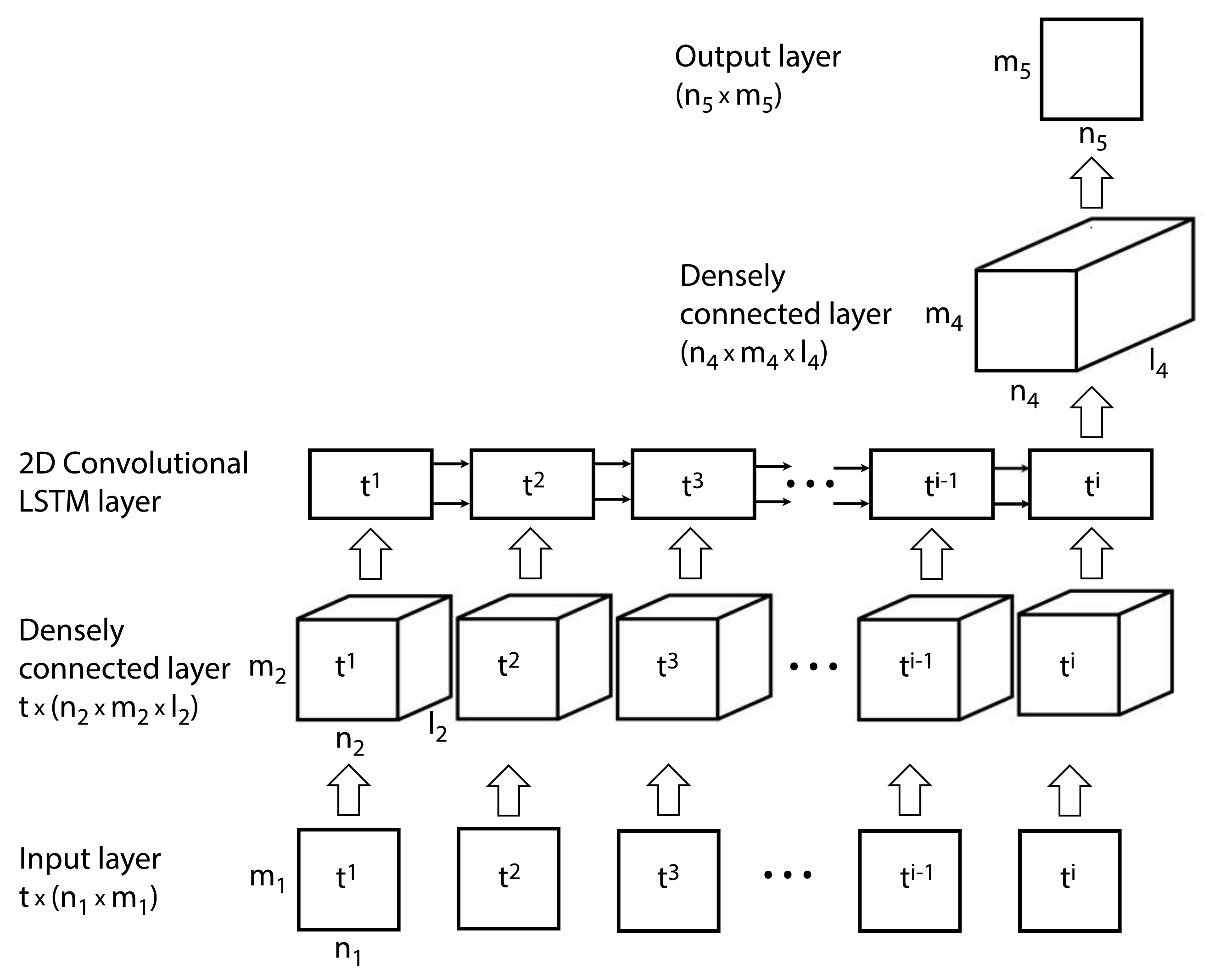}
\caption{Schematic view of the artificial neural network architecture. Give the input of a time-series of length $t^\mathrm{i}$ of 2D maps, passing it through a combination of densely connected layers and a layer with a series of LSTM cells including 2D convolutional operations, Deep-SANNE provides 2D output at the last time-step, $t^\mathrm{i}$.
}
\label{fig:NN_architecture}
\end{figure*}

The architecture of Deep-SANNE is illustrated in Fig~\ref{fig:NN_architecture}.
The input layer has $t^i = 24$ time steps, each of dimensions $n\times m = 30 \times 30$,
which is followed by a densely connected layer with a linear activation function and an increased number of channels, with a series of $t^i = 24$ batches of dimensions $n \times m \times l_1 = 30 \times 30 \times 64 $.
A densely connected layer connects each of its nodes to each of the nodes in the adjacent layers. The convolutional LSTM layer follows, with a series of $t^i = 24$ LSTM cells performing the temporal-convolutional operations 
at even more channels with dimensions $n \times m \times l_2 = 30 \times 30 \times 128$. 
The convolution kernel is 2D with $4\times 4$ cells, moving in the spatial $n,m$-plane. 
The rectified linear unit \citep[ReLU;][]{2010_relu} is used as activation function in the LSTM layer.
Only the output of the last LSTM cell is considered, which is fed to another densely connected layer with a linear activation function and dimensions $n \times m \times l_3 = 30 \times 30 \times 128$. 
Last, there is a final densely connected layer that samples down the dimensions to $n \times m = 30 \times 30$, which constitutes the output of the artificial neural network.
Further details on Deep-SANNE can be found at the git repository\footnote{\url{https://github.com/henrikeklund/Deep-SANNE/}}.

\subsection{3D MHD simulations for training and validation data}
\label{sect:methods,MHD}

Two dedicated numerical three-dimensional MHD Bifrost \citep{2011A&A...531A.154G} simulations of the solar atmosphere were used, one to train the neural network, and one to validate the estimates. \hea{These simulations were used as they} take nonlocal thermodynamic equilibrium (non-LTE) and nonequilibrium for the hydrogen ionization into account, which is important \hea{for studies} at millimeter wavelengths, where the majority of the radiation comes from the free-free emission. 

The training data were created from a re-run of a publicly available enhanced network (EN) simulation featuring the quiet Sun with two magnetic field regions of opposite polarity \citep{2016A&A...585A...4C}.
The re-run simulation has a cadence of 1~s, instead of the 10~s of the publicly available simulation, which matches the highest cadence offered so far for solar ALMA observations, and a total duration of approximately one hour. The simulation has $504$, $504$ and $496$ cells in the $x$-, $y$- and $z$-direction, respectively. The cell size is uniformly $48$~km, corresponding to $0.066$~arcsec in the horizontal ($x$, $y$) directions and varying between $19$~km and $100$~km in the vertical ($z$) direction, with $19$~km at the heights of the chromosphere and transition region, which are most relevant for the study of radiation at millimeter wavelengths.
\hea{This Bifrost simulation show intensities, scales, and oscillatory behavior at millimeter wavelengths that generally agree with observational data \citep{2015A&A...575A..15L, 2020A&A...635A..71W, 2021A&A...656A..68E,2021RSPTA.37900174J}, which makes it suitable for use as a model for training the neural network.}

The simulation that was used to create the validation data exhibits the quiet Sun without any large-scale magnetic fields with an average signed magnetic field strength of 5~G. It also represents the conditions of a coronal hole (CH). The simulation box has $1024$ cells with a cell size of $12$~km (corresponding to $0.0165$~arcsec) in the two horizontal directions ($x$, $y$) and $768$ cells with a cell size varying between $12-82$~km in the vertical direction ($z$).
This simulation has a 2~s cadence and a total length of 240~s.

\subsection{Preparation of observables}
\label{sect:methods,mm-observables}

The observables were calculated from the models using the radiative transfer code ART\footnote{\url{https://github.com/SolarAlma/ART)}} \citep{2021_art}, which solves the equation of radiative transfer along the vertical direction of the models for each column of cells across the field of view seen from the top. The code assumes formation of the radiation in LTE, but includes relevant sources of opacity in detail.
While the method presented in the current work is general and could be applied to handling observations at any wavelength, we focus here on calculations at wavelengths between $1.2$~mm - $1.3$~mm ($229$-$249$~GHz), corresponding to ALMA receiver band~6. 
The intensities were calculated at ten frequencies across band~6 \citep[see][for details]{2021A&A...656A..68E}.
The intensities ($I_\lambda$) were transformed to brightness temperatures ($T_\mathrm{b}$) through the Rayleigh-Jeans approximation at millimeter wavelengths \citep[see, e.g.,][]{2013tra..book.....W}.
The final maps of observables are represented by averages of the brightness temperature maps at the ten frequencies. A snapshot of the observables from the training simulation is shown in Fig.~\ref{fig:ori_conv_diff}a.

\hea{Deep-SANNE is currently trained to perform on science-ready data where possible effects of the sidelobes of the point spread function (PSF) have been taken care of.}
The brightness temperature maps are also downgraded to the angular resolution corresponding to observational data by convolution with a 2D kernel in the shape of the clean beam of the observations. The clean beam is a Gaussian fit to the central lobe of the \hea{PSF}.
The size and shape of the clean beam is calculated by doing synthetic ALMA observations with the simobserve task in the Common Astronomy Software Applications (CASA) for different receiver bands, interferometric array configurations, and positions of the target on the sky \citep[see][for details]{2021A&A...656A..68E}. 
The clean beam that is applied represents typical observations (the Sun standing high on the sky) with the highest resolution that is currently offered for solar observations at ALMA band~6, with the full width at half maximum (FWHM) along the minor and major axes is $0.69$~arcsec and $0.82$~arcsec, respectively (and a position angle of 80 degrees).

In order to avoid oversampling the image, it is common convention to chose an image cell size (pixel size) no smaller than approximately 3-5 times smaller than the FWHM of the clean beam when images are constructed from observations. 
For this reason, we regridded the brightness temperature maps with a cell size corresponding to $0.14$~arcsec to facilitate applications on already existing data sets.

\subsection{Training and validation of the neural network}\label{sec:methods-training_validation}

\begin{figure*}[tp!]
\includegraphics[width=\textwidth]{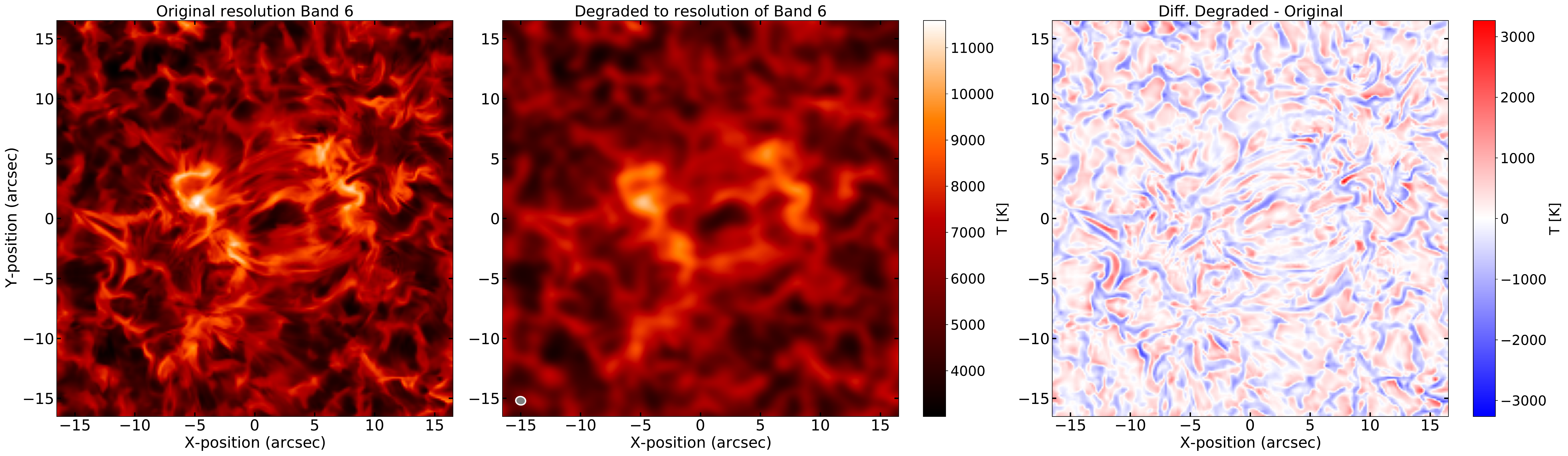}
\caption{Illustration of \hea{simulation} data used for training Deep-SANNE.
\textbf{a)} Synthetic brightness temperature map at $\lambda=1.25$~mm at the original high resolution, at $1800$\,s from the start of the \hea{training} simulation. 
\textbf{b)} The same brightness temperature map degraded to the angular resolution corresponding to ALMA observations. The clean beam with an FWHM of $0.69 \times 0.82$~arcsec is marked in the lower left corner. 
\textbf{c)} Difference between the degraded and the original map with absolute differences of up to $3200$~K.
}
\label{fig:ori_conv_diff}
\end{figure*}

For the training of the neural network, about $31000$ input sequences were sampled from the about one-hour time-series of brightness temperature maps at degraded angular resolution from the EN simulation (Fig.~\ref{fig:ori_conv_diff}b). 
Each input sample sequence consisted of 24 frames in total, with equidistant spacing every 10~s between t=0~s and t=60~s and every 2~s between t=68~s and t=100~s, and with a spatial extent of $30\times 30$~cells.
A spacing between the first frames of each input sample of 60~s in the temporal domain and 20 cells along each horizontal direction was used. Each sample was also rotated by $90$~degrees to further sample the data because the neural network is not rotational invariant.
Some of the input data were randomly excluded from the LSTM layer, with a dropout factor of $0.1$.
The neural network was trained to estimate the map of differences between the brightness temperatures at original and degraded resolution, relative to the degraded resolution (Fig.~\ref{fig:ori_conv_diff}c) following

\begin{equation}\label{Tbdiff_equation}
T_\text{b, difference} = \frac{T_\text{b, observational resolution} - T_\text{b, original high resolution}}{T_\text{b, observational resolution}}
.\end{equation}

To train the neural network, the optimization scheme Adam \citep{2014arXiv1412.6980K} was used
 together with a mean-squared error loss function and a learning rate of $1\times 10^{-4}$.
The neural network was allowed to learn for 15 epochs, and showed a smooth asymptotically declining loss function.
The accuracy of the estimates from the neural network was determined by feeding the neural network with the unseen brightness temperature maps from the CH simulation at degraded resolution.
As the neural network performs estimations in a small region at the time, the estimations need to be iterated across a larger field of view. 
Consequently, the neural network only needs to be applied to the spatial and temporal parts of the observations that is to be analyzed, which saves storage and computational costs in comparison to handling whole data sets.
The results of the validation are presented in Sect.~\ref{sec:results}.

\subsection{Observational data}\label{sec:methods-observational_data}
When Deep-SANNE was properly trained, it was also applied to perform estimations on observational ALMA data at $\lambda=1.25$~mm (receiver band~6), \hea{targeting a network region \citep{2021RSPTA.37900174J} at $(X,Y)=(-246,267)$~arcsec from the disk center. The data were} captured on April 22, 2017, between 15:59 UT and 16:43 UT (project ID: 2016.1.00050.S), with four calibration gaps of approximately $2$~min.
A high-cadence time series with $2$~s cadence was constructed with the Solar ALMA Pipeline (SoAP; Szydlarski et al., in prep.), including a multiscale CLEAN deconvolver algorithm \citep{2008ISTSP...2..793C, 1974A&AS...15..417H}, phase self-calibration, and combination of the interferometric data with total power (TP) measurements \hea{(see also, e.g., \cite{2020A&A...635A..71W, 2020A&A...644A.152E} for more details).}
The final science-ready observational data set, \hea{additional observational details, and information on co-observations} can be found in the Solar ALMA Science Archive\footnote{\url{http://sdc.uio.no/salsa/}} \citep[SALSA;][]{2022A&A...659A..31H}.

\section{Results} 
\label{sec:results}

\subsection{Brightness temperature correction factors, accuracy across the field of view, and refined image}

\begin{figure*}[tp!]
\includegraphics[width=\textwidth]{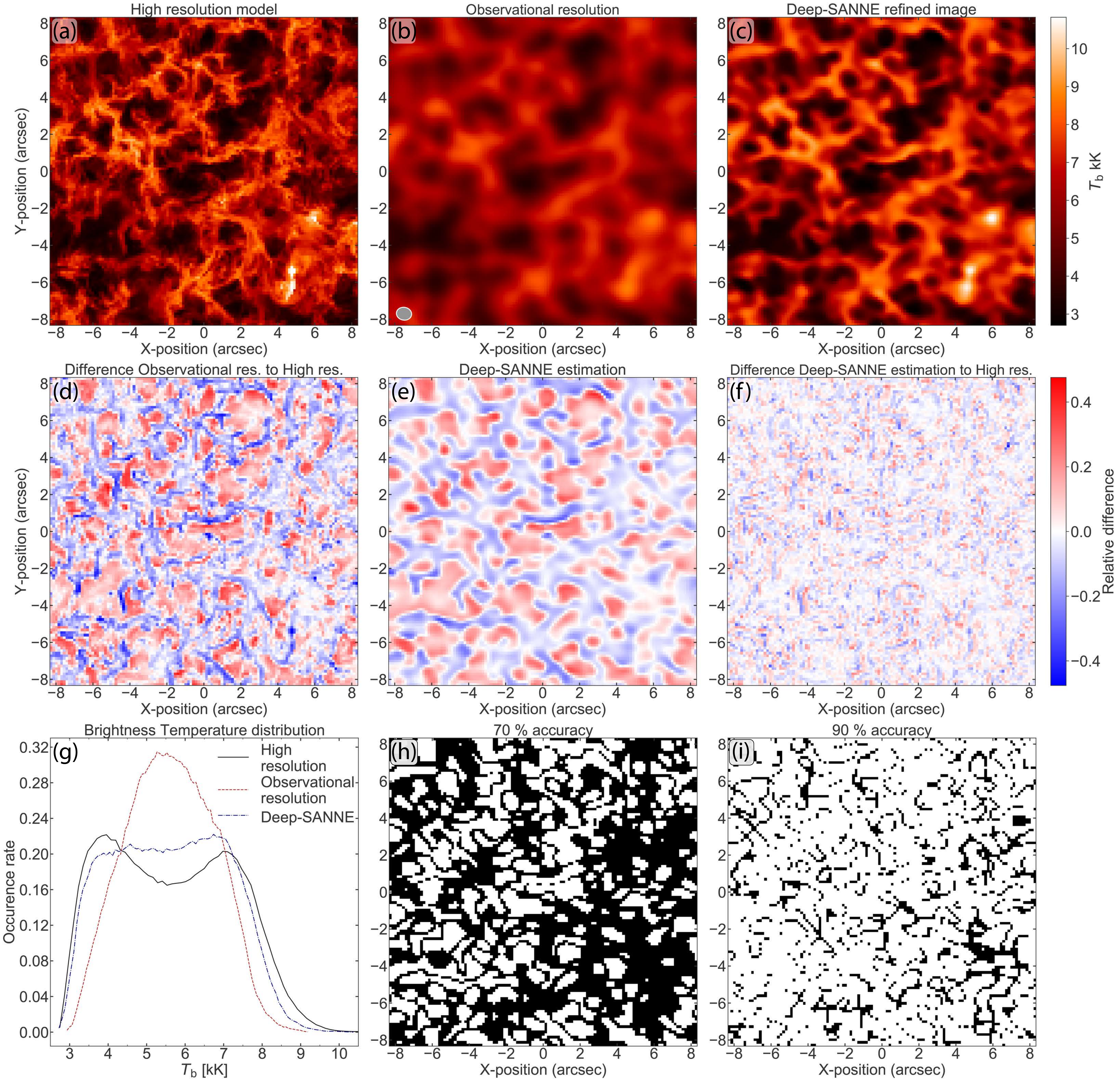}
\caption{Deep-SANNE estimates of the validation simulation. Panels (a), (b), and (c): Brightness temperatures across the field of view at $t=100$~s of the original model at high angular resolution,
the model degraded to the angular resolution corresponding to ALMA observations, with the clean beam resolution element marked in the lower left corner 
and the refined image from the Deep-SANNE estimation, respectively. The color map of the brightness temperatures is the same in each of the three panels.
(d) Differences between the maps of brightness temperature at the observational resolution and of the high-resolution model relative to the map at the observational resolution.
(e) Estimation of the difference map from Deep-SANNE.
(f) Difference map of the brightness temperatures of the Deep-SANNE refined image and the high-resolution model relative the Deep-SANNE refined image. The same color map is given in panels d--f.
(g) Distributions of the brightness temperatures between $t=100$~s--$240$~s of the highly resolved model, the observational resolution, and the Deep-SANNE estimate.
(h) and (i) Masks from Deep-SANNE showing the locations over the field of view at $t=100$~s (in black) where the accuracy of the brightness temperatures at observational resolution is least $70\%$ and $90\%$, respectively, compared to the highly resolved model.
}
\label{fig:CH_NN_validation}
\end{figure*}

In this section, we first check the performance of Deep-SANNE with the validation data set before applying it to real observations.
A map of the brightness temperatures at $1.25$~mm of the validation simulation (CH) is shown in Fig.~\ref{fig:CH_NN_validation}a for a snapshot at $t=100$~s, and the corresponding map degraded to the resolution of ALMA observations is given in Fig.~\ref{fig:CH_NN_validation}b. 

The relative difference between these two brightness temperature maps, following Eq.~(\ref{Tbdiff_equation}), is given in Fig.~\ref{fig:CH_NN_validation}d. 
The brightness temperatures of the data at the observational angular resolution deviate from those of the highly resolved map by an absolute relative factor of $\sim0.5$ or $\sim4750$~K. 
The Deep-SANNE estimate of the difference map over the FOV of the same snapshot ($t=100$~s) is shown in  Fig.~\ref{fig:CH_NN_validation}e. 
This estimate captures many of the small-scale features, also at scales smaller than the full width at half maximum of the clean beam
 (Sect.~\ref{sect:methods,mm-observables}). It also reveals the locations across the FOV that show under- or overestimated brightness temperatures at the observational resolution. 
These locations are indicated in Fig.~\ref{fig:CH_NN_validation}e in blue and red, and the magnitude of the deviation is indicated by the intensity of the color.

The estimated difference map can be multiplied with the observational data in order to produce a refined image of the brightness temperatures. 
The resulting Deep-SANNE refined image (at $t=100$~s) is shown in Fig.~\ref{fig:CH_NN_validation}c.
The figure clearly shows that Deep-SANNE manages to bring out accurate brightness temperatures and small-scale features that are not directly apparent in the map at the observational resolution. 

The median for the whole $140$~s time series of the absolute differences between the brightness temperatures of the observational data (Fig.~\ref{fig:CH_NN_validation}b) and the highly resolved data (Fig.~\ref{fig:CH_NN_validation}a) is $584$~K. The difference between the Deep-SANNE refined images (Fig.~\ref{fig:CH_NN_validation}c) and the highly resolved data (Fig.~\ref{fig:CH_NN_validation}a) alone is $290$~K.

Some features show a similar brightness temperature in the map at the observational resolution, which differs significantly at high resolution. 
Deep-SANNE distinguishes whether a feature at observational angular resolution corresponds to an extended event with a similar brightness temperature or to a combination of small brighter and small colder features, at high resolution. 
A few illustrating examples of this are described below.

(\textit{i}) The bright feature at (x,y)=(-0.5, -6.0)~arcsec, with a size of about $1.5\times1.5$~arcsec, shows a similar brightness temperature and size at the observational resolution (Fig.~\ref{fig:CH_NN_validation}b) as at the high resolution (Fig.~\ref{fig:CH_NN_validation}a). This is also reflected in the Deep-SANNE refined image (Fig.~\ref{fig:CH_NN_validation}c).

(\textit{ii}) The bright feature visible at (x,y)=(4.5, -6.5)~arcsec at the observational resolution originates from a combination of a much brighter but small feature in the direct vicinity of the dark features at high resolution. The small bright features are indicated in the Deep-SANNE refined image.

(\textit{iii}) The bright feature at (x,y)=(-4.0, 1.5)~arcsec shows a roughly uniform brightness temperature at the observational resolution, which originates from the combination of a small ($\sim0.5$~arcsec) relatively dark region surrounded by a ring of brighter features at high resolution. This complex structure is captured in the Deep-SANNE refined image. 
When performing an in-depth analysis of such a bright feature in observational data, it might be tempting to sample the brightness temperature in the center of the feature, which in this particular case is the least accurate position.

We also calculated the differences of the brightness temperatures of the refined image (Fig.~\ref{fig:CH_NN_validation}c) from the image at high angular resolution (Fig.~\ref{fig:CH_NN_validation}a) relative to the refined image, which results in the map that is shown in Fig.~\ref{fig:CH_NN_validation}f.
In this figure, the uncertainty of the Deep-SANNE refined image is indicated, the uncertainties with about arcsecond scales visible in Fig.~\ref{fig:CH_NN_validation}d are removed, and only structures of a small spatial size of about some pixels are visible.
Uncertainties on scales much smaller than the clean beam are expected, and this indicates that the neural network is not over fitted.
However, there might be small potential improvements on this in the image gridding, which is discussed in Sect.~\ref{sect:disc-image_gridding}. 
The average absolute value of the relative differences is $0.07$ (both at t=$100$~s, Fig.~\ref{fig:CH_NN_validation}f, and for the whole time series), and the estimated brightness temperatures of the Deep-SANNE refined image \ref{fig:CH_NN_validation}c) are thus accurate at $93\%$ on average.

The distributions of the brightness temperatures for the whole time series at high resolution, observational resolution, and of the Deep-SANNE refined images are given in Fig.~\ref{fig:CH_NN_validation}g. 
The shape of the distribution of brightness temperatures varies with the magnetic field strength and topology of the region, and therefore gives valuable information for the interpretation of the data.
There is a bimodal distribution at high resolution (Fig.~\ref{fig:CH_NN_validation}g), with one peak around $4$~kK and one around $7$~kK, which is expected for a quiet-Sun region \citep{2021A&A...656A..68E}.
At observational angular resolution, the double-peak feature is not visible, and the distribution instead shows a single more narrow peak centered around the average value of $5590$~K.
This value agrees well with the average of the synthetic observables at $1.25$~mm over the quiet-Sun region in the EN simulation (training simulation; Fig.~\ref{fig:ori_conv_diff}) of $5254$~K \citep{2021A&A...656A..68E}, but is lower than what is reported for observational quiet-Sun data at $1.25$~mm, for instance, $5957$~K \citep{2021RSPTA.37900174J} and $6150$~K \citep{2017SoPh..292...88W}. The observations most likely show higher average temperatures because they also partly include components of overlying magnetic fields \citep{2021RSPTA.37900174J}, which is not present in this coronal hole simulation (Fig.~\ref{fig:CH_NN_validation}) or in the corners of the EN simulation.
The refined images from Deep-SANNE restore the signature of the bimodal distribution of the brightness temperatures, even though the distribution is weaker than at the original high resolution. 
The estimates are conservative, with a brightness temperature span within that of the highly resolved model. 
The kurtosis of the distribution of the brightness temperatures at the high-resolution model is $-1.00,$ and at the observational resolution, it is only $-0.66$, while at the Deep-SANNE estimate, it is $-0.97$. 
There is thus an improvement of the deviation from the highly resolved kurtosis from $34\%$ at the observational resolution to only $3\%$ using Deep-SANNE.

Because the accuracy of the brightness temperatures at observational resolution varies much at small scales across the FOV, the choice of locations when performing in-depth analyses of features is important. 
Masks that indicate where in the field of view the brightness temperatures are probably to be most accurate can be created from the Deep-SANNE output. Locations with a brightness temperature that is $70\%$ and $90\%$ accurate at least in the observational resolution are indicated,  in Fig.~\ref{fig:CH_NN_validation}h--i, to an accuracy of the positions of $90\%$ at least.
The temperatures of the most reliable locations span a wide range. 
There is no simple relation between the temperature and the reliability in the case of the complex structures of the highly dynamic solar atmosphere, and there are locations with both low and high temperatures with high accuracy.

\subsection{Estimations of dynamic brightening events}\label{sect:results-Tb_events}

\begin{figure*}[t]
\centering
\includegraphics[width=\textwidth]{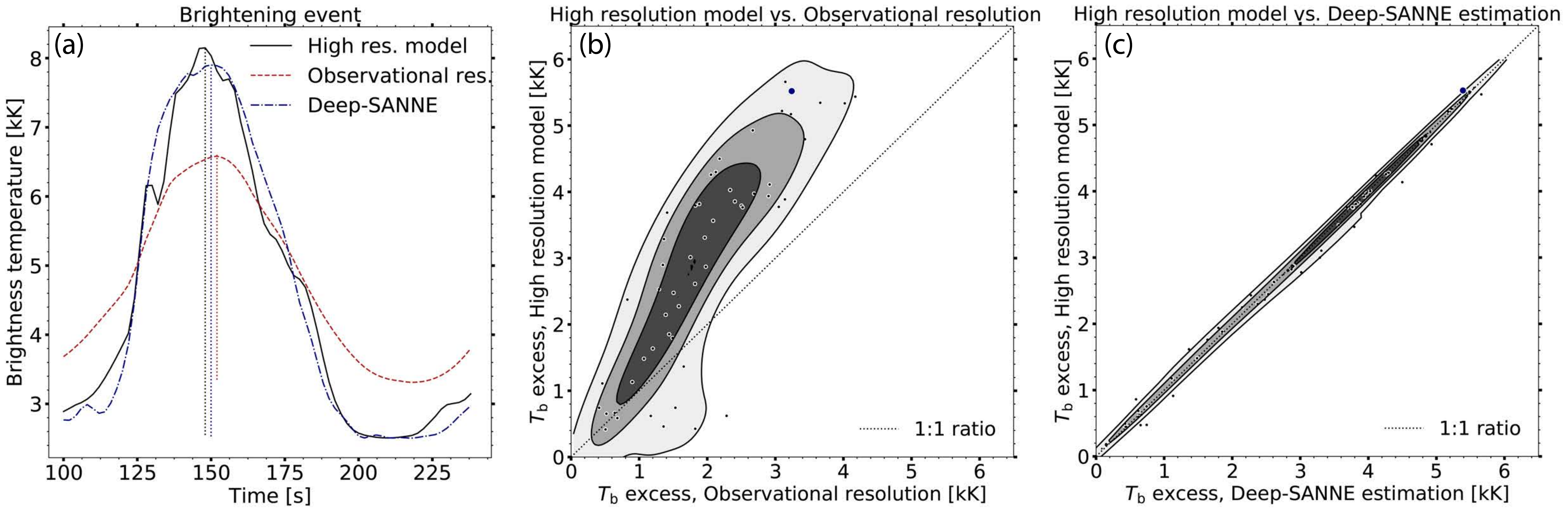}
\caption{Degradation of brightening events \hea{in the simulation}. (a) Example of a brightening event with the temporal evolution of the brightness temperature at the degraded angular resolution corresponding to observations (dashed red), of the original high-resolution model (solid black), and of the Deep-SANNE refined images (dot-dashed blue). The apparent magnitude of the temperature excess is in each case marked by the vertical dotted lines.
(b) Density plot of the magnitudes of brightening events at high resolution against the observational resolution.
(c) Density plot of the magnitudes of brightening events at high resolution against the estimate from the neural network.
All events are indicated by circular markers on top of the density plots, with the example from panle (a) given in blue for reference.
}
\label{fig:tb_events}
\end{figure*}

\begin{figure*}[t]
\centering
\sidecaption
\includegraphics[width=12cm]{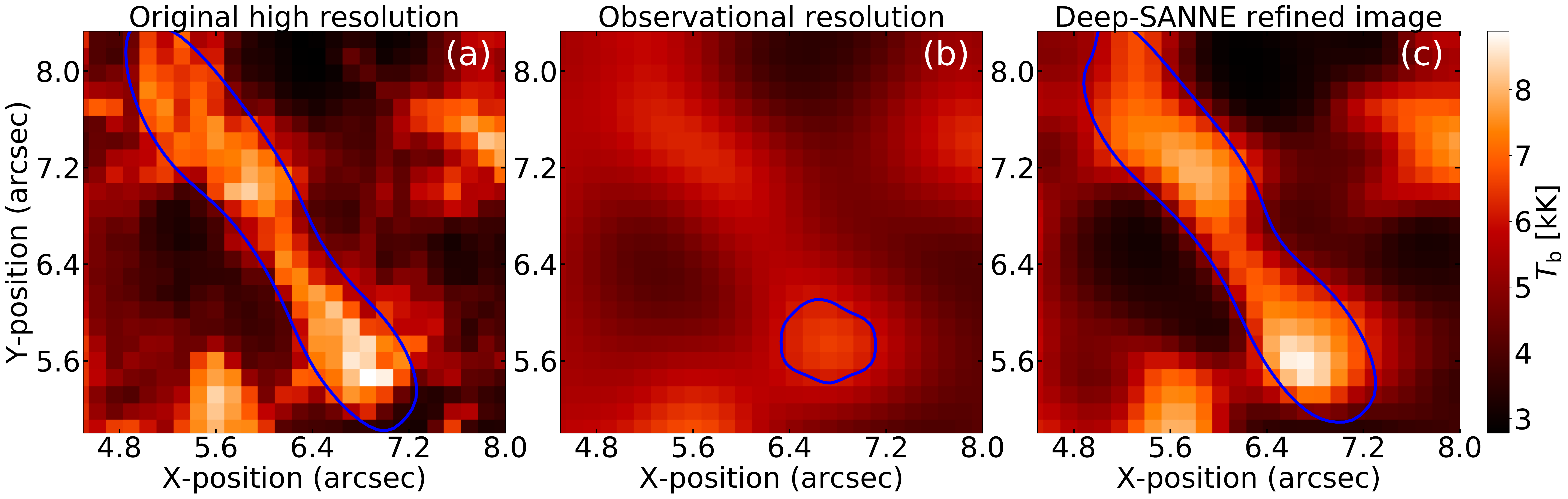}
\caption{Zoom-in of a brightening event in the simulation. \hea{(a)} At original high resolution, \hea{(b)} at a resolution corresponding to observational data, and \hea{(c)} the Deep-SANNE refined images. The blue contour marks the spatial size of the detected event at the respective series of images. 
}
\label{fig:event_size}
\end{figure*}

\cite{2021A&A...656A..68E} showed that brightening events showing an excess of temperature are degraded as a result of the limited spatial resolution of the observations, which limits the interpretation and analysis of these events.
The Deep-SANNE refined images can be used to acquire more accurate data on the brightness temperature excess and the size and shape of the brightening events.

The validation data set was searched for brightening events by searching for peaks in the temporal evolution of the brightness temperature at each position in the field of view. The peaks that were found were grouped into individual brightening events by a k-means clustering technique \citep{Everitt:1972aa}. 
The event-detection algorithm is described by \cite{2020A&A...644A.152E}, where the specific details can be found.
In total 53 separate brightening events were detected.

An example of the temporal evolution of a brightening event is shown in Fig.~\ref{fig:tb_events}a at a degraded resolution, at the original high resolution, and at the Deep-SANNE estimate.
When the temporal domain is used by adding the RNN module to the CNN (Sect.~\ref{sect:methods-deep_sanne}), a smooth continuous behavior of the Deep-SANNE estimate of the brightening event is obtained that closely follows the main profile of the high-resolution model. 
This specific event shows an excess brightness temperature of about $5.52$~kK at high resolution, but only $3.25$~kK at observational resolution, corresponding to only $58.9\%$ to that of the high resolution. 
However, in the refined images of Deep-SANNE, the event shows an excess brightness temperature of $5.39$~kK, amounting to $97.6\%$ of the high-resolution value.

The excess brightness temperatures of each event at high resolution compared to that at observational resolution are given in Fig.~\ref{fig:tb_events}b.
Most of the events are underestimated, showing lower brightness temperature excess at the observational resolution than at high resolution. 
There are also a few weaker events that are overestimated. They lie below the one-to-one ratio. This might be because they are in the close vicinity to a bright event. 
However, some events are better resolved and more accurately represented at the observational resolution than others, and even though the distribution is fairly linear and could be roughly described by a linear function as in \cite{2021A&A...656A..68E}, the distribution shows a wide spread and gives rise to significant uncertainties.

The equivalent plot comparing the excess brightness temperatures of the events at high resolution to the Deep-SANNE images is given in Fig.~\ref{fig:tb_events}c.
Deep-SANNE efficiently identifies how well resolved the individual events are, which is indicated by the alignment of the events along the one-to-one ratio in Fig~\ref{fig:tb_events}c.
The neural network corrects well for the degradation of the events, regardless of whether they are under- or overestimated in the . 
The brightening events show magnitudes with an average accuracy of only $43.7\%$ at the observational resolution compared to at high resolution, while Deep-SANNE estimates the magnitudes of the brightening events to an average accuracy of $94.0\%$.

The spatial size and shape of a brightening event \hea{in the simulation} is shown in Fig.~\ref{fig:event_size}. 
The event \hea{is elongated with a length of about} $4$~arcsec \hea{and a width of about $1$~arcsec} at high resolution (Fig.~\ref{fig:event_size}a). At the observational resolution, the event only extends about $0.8$~arcsec \hea{in each direction} (Fig.~\ref{fig:event_size}b), where the top parts do not show a sufficiently high brightness temperature excess as a result of the spatial averaging with the surrounding dark features. They are therefore overlooked by the detection algorithm, which needs to be conservative to avoid sampling noise in the data in the case of observations.
In the Deep-SANNE refined images (Fig.~\ref{fig:event_size}c), the event shows a \hea{length of about} $3.8$~\hea{arcsec} and a similar shape as in the high resolution.
At low resolution, features tend to show more simple shapes with less complex structures, and it is relatively easy to define borders.
However,  as a result of the complex filament structure of the brightening events, it becomes increasingly demanding with higher resolution to set contrasts and thresholds to define the extent of events and distinguishing interfering events from each other.  
From the events detected here, there is an improvement of the accuracy of the sizes of the brightening events from $49\%$ at observational resolution to at least $78\%$ using the Deep-SANNE refined images, compared to at the highly resolved images. 
\hea{The brightening feature in Fig.~\ref{fig:event_size} appears to be slightly wider, but still comparable to the angular resolution, which resulted in the degradation of the intensity that Deep-SANNE is able to restore reasonably well. 

Using ALMA 3.0~mm data, \cite{2017ApJ...841L...5S} and \cite{2019ApJ...875..163R} reported studies of a bright plasmoid ejection event, and \cite{2021A&A...651A...6B} reported an analysis of a number of brightening events. Most of these events show a spatial size that is comparable to the angular resolution, which could indicate that their signatures could be significantly degraded. 
The plasmoid ejection event shows an increase in the brightness temperature of up to 220~K \citep{2019ApJ...875..163R}, which agrees with many of the brightening events detected in the Deep-SANNE training data at degraded resolution \citep{2021A&A...656A..68E} and the events in the lower left corner of Figs.~\ref{fig:event_size}b--c. 
The method of Deep-SANNE could thus be applied to these type of brightening events to acquire refined temperatures and shapes for a potentially improved scientific analysis.
}

\subsection{Application of Deep-SANNE on observational ALMA data}\label{sect:results-obs_data}

\begin{figure*}[t]
\centering
\includegraphics[width=\textwidth]{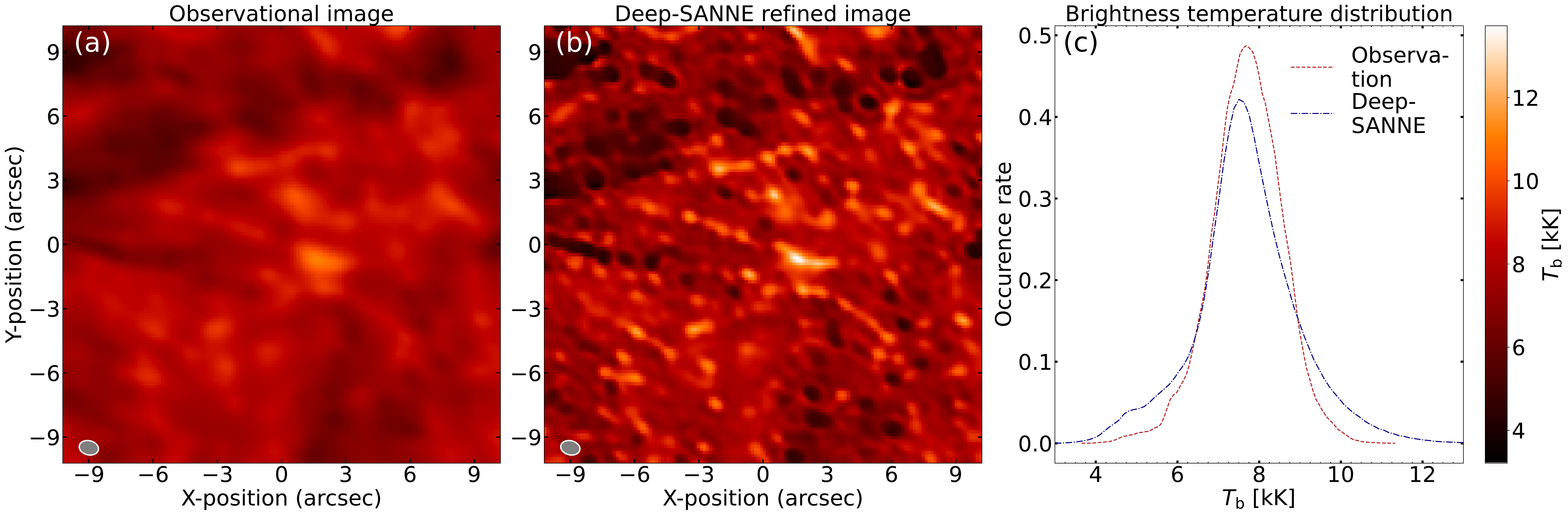}
\caption{\hea{Observational ALMA data at $1.25$~mm from 2017-04-22} (a) Snapshot image from ALMA observations at 16:01:37.
(b) Corresponding Deep-SANNE refined image.
(c) Brightness temperature distributions of the \hea{time-series of} observations and the Deep-SANNE refined images.
}
\label{fig:obs_data}
\end{figure*}

A snapshot of observational ALMA data \hea{at $\lambda=1.25$~mm (see Sect.~\ref{sec:methods-observational_data})} at $t=150$~s from the beginning of the data set is shown in Fig.~\ref{fig:obs_data}a, and the Deep-SANNE refined image for the same time step 
is shown in Fig.~\ref{fig:obs_data}b.
The refined image shows significantly improved contrast and allows a more meaningful analysis of the small-scale features. 

The dot-like structure dominating the features in the refined image originates from the techniques used to reconstruct the observational image. The features are imprints of the clean beam, which can be seen by comparison to the size, shape, and orientation of the clean beam, for which the FWHM is shown in the lower left corner of the images, which we discuss further in Sect.~\ref{sec:disc}. 
In this sense, Deep-SANNE performs well at also revealing potential artifacts that may have been introduced to the images during the reconstruction process, and can thus be used to assist sampling the real signals in a meaningful way.
From the refined image, Fig.~\ref{fig:obs_data}b, it becomes clear that the choice of locations for an in-depth analysis and eventual averaging techniques across a feature is important when studying the intensities, shapes, evolution, and potentially oscillatory behavior of small-scale features.

The distributions of brightness temperatures across the FOV between $t=100$ -- $376$~s for the observations and the Deep-SANNE refined images are given in Fig.~\ref{fig:obs_data}c.
\hea{The average brightness temperature is $7664$~K in both the observational data and the Deep-SANNE refined images. The average temperature is thus well preserved by Deep-SANNE, but the}
brightness temperature distribution of \hea{the refined} images represents cold and hot features better, as expected at a less degraded resolution \citep{2021A&A...656A..68E}.
This observation targets a network region with a combination of vertical, near-vertical, and horizontal magnetic fields in the FOV \citep{2021RSPTA.37900174J}.
\hea{The} single peaked brightness temperature distribution agrees very well with the distribution that is seen for the magnetically enhanced central parts of the of the field of view of the training data set (EN) \citep[cf. Fig.4 of][]{2021A&A...656A..68E}. 
The average brightness temperature of these observational data (Fig.~\ref{fig:obs_data}c) agrees with the average value of $7746$~K reported by \cite{2021RSPTA.37900174J} for another observation at $1.25$~mm toward a target showing similar magnetic field conditions, and of $7417$~K for the $1.25$~mm synthetic observables of the network region in the EN simulation (the training simulation) reported by \cite{2021A&A...656A..68E}.


\section{Discussion}
\label{sec:disc}

\subsection{Parameter space of the MHD models used for the training}

The results presented here are based on a training of the artificial neural network by using the radiative transfer calculations on a simulation of approximately 3400~s and $33\times33$~arcsec and on the validation of the simulation of merely 140~s and $15\times 15$~arcsec.
The different simulations have different physical setups, and if the neural network were trained at both simulations and even more additional simulations, it might be able to work even better to perform estimations of observational data. 
\hea{For reliable results, Deep-SANNE should be applied for studying features that are represented within the parameter space of the model(s) used for the training.
The current version of Deep-SANNE can thus be applied to quiet-Sun or network patches, as this is what the training model features (Sect.~\ref{sect:methods,MHD}),
and the outputs could in principle be used to study any type of feature, dynamic event, or oscillations. 
A higher accuracy of the spatial distribution of the brightness temperatures also contributes to a more accurate analysis of the temporal evolution of dynamic events and oscillations in the data \citep{2021A&A...656A..68E}. 
Applications or analyses should be performed with caution on events that show brightness temperature contrasts or a dynamic evolution that is very different than what is represented in the training model. 
}

The training of Deep-SANNE could be extended by including simulations showing other properties of small-scale dynamics, beyond what is represented in the current training simulation (Sect.~\ref{sect:methods,MHD}). \hea{This might be applicable to sunspots, for example, given adequate simulations of such regions.}

\subsection{Image gridding}\label{sect:disc-image_gridding}

The resolution of observations is determined by the diffraction pattern of the PSF (or the clean beam) and the minimum angular scale required to distinguish two point sources.
The cell size of the image gridding (pixel size) is mostly set in relation to the size of the clean beam, 
and based on studies of static relatively targets, it is commonly
set to 
between three to five times smaller than the FWHM of the clean beam. However, when targets with strong dynamic features are observed,
the temporal domain provides useful information for constructing the final images, and the classical approach of image gridding could be challenged.
Nonetheless, in the current work, Deep-SANNE performs estimations on maps of mm wavelength observables (1.25~mm)
with a cell resolution corresponding to $0.14$~arcsec,
(Sect.~\ref{sect:methods,MHD}). This is 4.9-5.9 times smaller than the clean beam major and minor axes (Sect.~\ref{sect:methods,mm-observables}).
While it is interesting to investigate an optimal cell size for observations of the solar atmosphere when using a temporal-spatial technique such as Deep-SANNE, 
the estimations are performed on the typical grid resolution that was used to reconstruct current observational data so far \citep{2022A&A...659A..31H}. Deep-SANNE can therefore be directly applied on existing science-ready data sets.

\subsection{Properties of the observational data}

Deep-SANNE expects time series with a cadence of 2~s or higher. 
Further studies need to be made to establish the dependence on the cadence and results of applications to data sets with lower cadence.

The currently highest angular resolution offered for solar ALMA observations is down to $0.62$~arcsec \citep{ALMA_Cycle_9_Prop_Guide} at band~6 ($1.25$~mm). However, achieving this resolution isotropically is limited to optimal conditions with the Sun close to zenith.
For an interferometric array such as ALMA, the location of the target on the sky influences the shape of the PSF and the corresponding clean beam. 
When the target stands lower in the sky, the clean beam becomes more eccentric. 
In the current work, applications were made
with a slightly eccentric clean beam based on simulated observations corresponding to normal conditions similar to what is achieved in the ALMA observations.
The same method can be used for data at other wavelengths, even at other wavelength regimes, and at other angular resolutions.
The angular resolution is dependent on the wavelength, but also on the size of the aperture, which in the case of an interferometric array is determined by the antenna positions.
The neural network can be trained to recognize the skewing that a more eccentric clean beam of a specific data set inflicts on the data and the limitations of detecting the underlying dynamical features.

Extending the current work by applying the neural network method on data at lower angular resolution, at lower cadence, or with a very eccentric clean beam would be useful to study how these parameters effect the observability of small-scale features.
Furthermore, the observations used here were reconstructed over the full receiver band 6 (1.2~mm~-~1.3~mm; Sect.\ref{sect:methods,mm-observables}), but Deep-SANNE might also be applied to images constructed from the receiver sub-bands individually at 1.2~mm and 1.3~mm \citep{ALMA_Cycle_9_Prop_Guide}, allowing a mapping of the slope of the brightness temperature continuum and gaining additional information for the analysis of the small-scale features \citep{2022arXiv221105586E}.


\subsection{Identification of noise in the observational images}

\begin{figure*}[tp!]
\centering
\includegraphics[width=\textwidth]{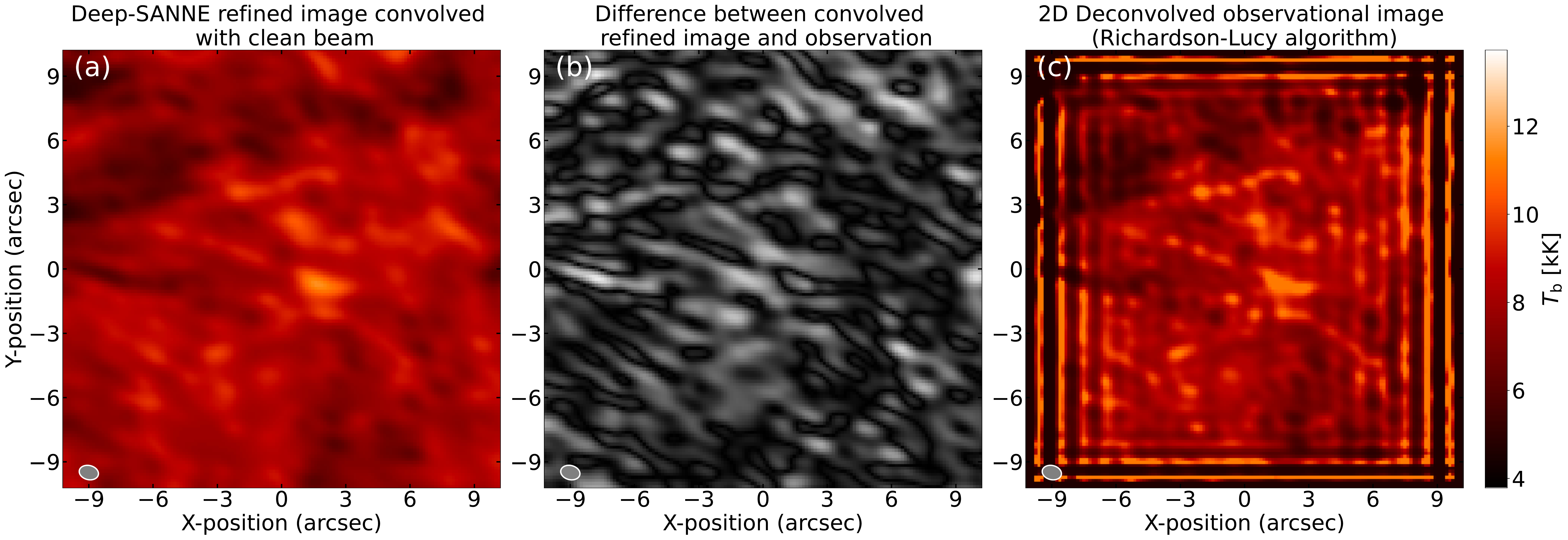}
\caption{Comparison of Deep-SANNE refined image. (a) Deep-SANNE refined image of the observational data (shown in Fig.~\ref{fig:obs_data}b) convolved with the clean beam corresponding to the observation. The FWHM of the clean beam is marked in the lower left corner.
(b) Difference between the convolved Deep-SANNE refined image (panel a) and the observational image (Fig.~\ref{fig:obs_data}a).
(c) Observational image (Fig.~\ref{fig:obs_data}a) deconvolved with the clean beam in the 2D spatial domain, using a standard Richardson-Lucy algorithm.
}
\label{fig:obs_data_convolution}
\end{figure*}
A significant fraction of the training data was randomly excluded in the training process of the neural network (Sect.\ref{sec:methods-training_validation}). This helps it to perform well on observations with incomplete data.

As mentioned above (Sect.~\ref{sect:results-obs_data}), to remove the impact of the side-beams of the PSF, the CLEAN deconvolution process \citep{1974A&AS...15..417H} was applied to the observational data (Fig.~\ref{fig:obs_data}a), which is the common case for an analysis of solar ALMA data published so far.
In short, CLEAN is an iterative process subtracting fractions of the PSF from the data (dirty image), determined by a gain factor and adding the corresponding clean beam to a model image (clean image) of the target until a pre-defined threshold is reached, resulting in a final model image composed of a collection of clean beams.
This method is optimal for sparsely distributed point sources and gives rise to certain imaging artifacts for extended targets, such as the Sun. 
The magnitudes and contrasts of these artifacts are not only dependent on the spatial distribution of the intensities and contrasts of the target, but also largely depend on the specific parameters used in the iterative process.
The final observational (model) image consists of a collection of clean beams, which is more clearly revealed in the high contrast of the Deep-SANNE refined image (Fig.~\ref{fig:obs_data}b).

Taking the Deep-SANNE refined image  (Fig.~\ref{fig:obs_data}b) and convolving it with the clean beam of the observational data as a 2D kernel, results in the image shown in Fig.~\ref{fig:obs_data_convolution}a. The small-scale features in this convolved image are remarkably similar to those in the observational image (Fig.~\ref{fig:obs_data}a), but the contrast is higher.
A map of the differences between the convolved image and the observational image is displayed in Fig.~\ref{fig:obs_data_convolution}b.
The difference map shows elongated features that appear smooth and continuous over several arcseconds and are not dominated by the shape of the clean beam, as apparent in Fig.~\ref{fig:obs_data}b). This shows that the neural network does not cause the dot-like structures in general.
There are some features in the difference map (Fig.~\ref{fig:obs_data_convolution}b) at the shape of the clean beam, corresponding to the features that the neural network can determine being under-resolved, however.

The result of applying a standard Richardson-Lucy 2D deconvolver algorithm \citep[with 350 iterations;][]{1972JOSA...62...55R, 1974AJ.....79..745L} to the observational image (Fig.~\ref{fig:obs_data}a) is shown in Fig.~\ref{fig:obs_data_convolution}c. 
The fringes are caused by edge effects from the deconvolution process with the relatively large clean beam 2D kernel. To be conservative with the brightness temperatures from the observational data, no edge preserving techniques were applied here, and therefore only a rather small central part of the image is useful for a reliable scientific analysis.
The 2D convolution algorithm improves the contrasts of the image, also revealing the dot-like structures, but it is unable to distinguish how well-resolved different features are.
However, by comparing the 2D deconvolution output (Fig.~\ref{fig:obs_data_convolution}c) to that of the Deep-SANNE refined image (Fig.~\ref{fig:obs_data}b),
the added information coming from the temporal evolution that gives Deep-SANNE the ability to distinguish the features that are under-resolved from the better resolved features becomes clear. 
For example, there is the somewhat extended bright feature in the centre FOV, (around $x,y$)=(2\arcsec,-1\arcsec), with a nearly uniform brightness temperature in the observational image (Fig.~\ref{fig:obs_data}a) that in the 2D deconvolved image (Fig.~\ref{fig:obs_data_convolution}c) appears almost equally uniform, while Deep-SANNE manages to recognize from the temporal dynamics that this feature is composed of a combination of a less bright component and a smaller bright point in the centre, similarly to what is seen in the results of the simulations (Sect.~\ref{sec:results}).

Deep-SANNE thus works as a deconvolver, effectively revealing the information of the features in the observational image with improved contrasts. Because it is based on both temporal and spatial dynamic variations of the brightness temperature, it works better than a standard 2D kernel deconvolver in estimating contrast corrections of the small-scale features.

The CLEAN algorithm serves as a very good first approximation to make use of these excellent solar observations at mm wavelengths. 
It is however very important to be aware of the limitations of the finalized science-ready data sets when they are analyzed.
If the data contain artifacts that are introduced in the imaging process, it is important to reveal them so that they can be taken into account as uncertainties. In this aspect, the Deep-SANNE refined images provide valuable tools for meaningful scientific interpretations.


\section{Summary and conclusions} 
\label{sec:conc}

We presented Deep-SANNE, which performs estimations of 
the differences of brightness temperatures at a high resolution and that seen at observations with limited angular resolution.

The deep neural network was trained on high-cadence time-series of maps of synthetic observables from a 3D MHD Bifrost simulation that shows a significantly large span of physical parameters for small-scale dynamics with a mixture of quiet-Sun and magnetically enhanced conditions.
Deep-SANNE learns from a large statistical basis of the temporal evolution of small-scale spatial features and dynamics and from the intensity contrast degradation that comes with limited angular resolution.
In observations at limited angular resolution, Deep-SANNE recognizes patterns of the small-scale dynamics in the data and estimates the intensities of the underlying features at high resolution. 
Deep-SANNE functions as a deconvolver technique, using information of dynamics in both the spatial and temporal domains to give the distribution of brightness temperatures at increased contrast of a refined angular resolution. This helps to identify features that otherwise might easily be overlooked.
Including the temporal domain allows distinguishing how well resolved different features are and results in better estimates of the contrast degradation than a standard 2D deconvolver.
The performance of Deep-SANNE was validated using another Bifrost simulation, featuring quiet-Sun conditions with a different magnetic field topology. This simulation was not previously seen by the neural network. 

The method shown in the current work could be applied to observations at various wavelengths and angular resolutions, but the applications of the current work are focused on solar ALMA observations with the setup for the highest resolution offered so far, at $1.25$~mm (receiver band~6) with a resolution between $0.69$--$0.82$~arcsec.
The relatively high resolution for mm wavelengths is favorable for resolving the small-scale features in the data, but the intensity contrast of the small-scale features are still significantly degraded. 
At lower angular resolution, the degradation of contrasts would be even more significant and the method of the current work might be more valuable. 
Deep-SANNE provides maps of correction factors that indicate the accuracy of the brightness temperatures of the observations across the field of view.
These maps can be used to create masks that only show locations of the brightness temperature with high accuracy, which is useful for selecting locations for an in-depth analysis of the data.
The refined images from Deep-SANNE can be used for studying all types of dynamic events, spatial sizes, and shapes of bright and dark features, their temporal evolution, and oscillatory behavior in the data.

In the current work, applications on brightening events were made to show the large improvements of using Deep-SANNE to estimate their excess brightness temperature, size, and shape.
Deep-SANNE can recognize whether a feature in the observational data is well resolved and performs well in estimating the spatial size, shape, and intensity of both dark and bright features.
That is, for an example, if an extended diffuse feature seen at degraded resolution corresponds to an extended diffuse feature or to a small point-like feature at high resolution.
The Deep-SANNE estimates were applied to correct for the degradation of chromospheric brightening events. The average accuracy of the magnitude of a brightening event shown at high resolution increased from $43.7\%$ at the observational resolution to $94\%$ at the Deep-SANNE refined images.

In addition to the significantly improvement of the intensity contrast of small-scale features, the Deep-SANNE refined images also help to reveal potential artifacts that were introduced to the data in \hea{the image} reconstruction process. This is important to include in the scientific analysis.

The estimates and refined images from Deep-SANNE allow performing a more precise analysis on any application where the variances and dynamic evolution of intensities, sizes, and shapes of spatial features are important. This allows a more meaningful interpretation of the observational data.


\section*{Acknowledgments}
The author wish to thank 
Miko\l{}aj Szydlarsk for running the Bifrost simulations,
Reyna Ramirez de la Torre for valuable input and support on the development of the project and writing the manuscript, Souvik Bose and Jayant Joshi for motivating discussions on the topic, and Carlos José Díaz Baso, Jaime de la Cruz Rodríguez and Jorrit Leenaarts for valuable input on the final stages of the manuscript.
This work was supported through the CHROMATIC project (2016.0019) funded by the Knut and Alice Wallenberg foundation, by the SolarALMA project, which has received funding from the European Research Council (ERC) under the European Union’s Horizon 2020 research and innovation programme (grant agreement No. 682462) and
the Research Council of Norway through its Centres of Excellence scheme, project number 262622, and through grants of computing time from the Programme for Supercomputing.
This paper makes use of the following ALMA data: ADS/JAO.ALMA$\#$2016.1.00050.S. ALMA is a partnership of ESO (representing its member states), NSF (USA) and NINS (Japan), together with NRC(Canada), MOST and ASIAA (Taiwan), and KASI (Republic of Korea), in co-operation with the Republic of Chile. The Joint ALMA Observatory is operated by ESO, AUI/NRAO and NAOJ. We are grateful to the many colleagues who contributed to developing the solar observing modes for ALMA and for support from the ALMA Regional Centres.

\bibliographystyle{aa}
\bibliography{ms}

\end{document}